\documentclass[prl,10pt,twocolumn,unsortedaddress]{revtex4-1}

\usepackage{graphicx}

\newcommand{\ket}[1]{|{#1}\rangle}
\newcommand{\vis}{{\mathcal V}}

\begin{document}

\title{Longer-Baseline Telescopes Using Quantum Repeaters}

\author{Daniel Gottesman}
\email{dgottesman@perimeterinstitute.ca}
\affiliation{Perimeter Institute for Theoretical Physics, Waterloo, Ontario, Canada}
\author{Thomas Jennewein}
\email{thomas.jennewein@uwaterloo.ca}
\affiliation{Institute for Quantum Computing, University of Waterloo, Waterloo, Ontario, Canada}
\author{Sarah Croke}
\email{scroke@perimeterinstitute.ca}
\affiliation{Perimeter Institute for Theoretical Physics, Waterloo, Ontario, Canada}

%\date{}

\begin{abstract}
We present an approach to building interferometric telescopes using ideas of quantum information.  Current optical interferometers have limited baseline lengths, and thus limited resolution, because of noise and loss of signal due to the transmission of photons between the telescopes.  The technology of quantum repeaters has the potential to eliminate this limit, allowing in principle interferometers with arbitrarily long baselines.
\end{abstract}


\maketitle

The two primary goals for a telescope are sensitivity and angular resolution.  Interferometry among telescope arrays has become a standard technique in astronomy, allowing greater resolving power than would be available to a single telescope.
In today's IR and optical interferometric arrays~\cite{Monnier,Michelson}, photons arriving at different telescopes must be physically brought together for the interference measurement, limiting baselines to a few hundred meters at most because of phase fluctuations and photon loss in the transmission.  Improved resolution would, if accompanied by adequate sensitivity, have many scientific applications, such as detailed observational studies of active galactic nuclei, more sensitive parallax measurements to improve our knowledge of stellar distances, or imaging of extra-solar planets.

The field of quantum information has extensively studied the task of reliably sending quantum states over imperfect communications channels.  The technology of quantum repeaters~\cite{Repeaters} can, in principle, allow the transmission of quantum states over arbitrarily long distances with minimal error.  Here we show how to apply quantum repeaters to the task of optical and infrared interferometry to allow telescope arrays with much longer baselines than existing facilities.  The traditional intended application for quantum repeaters is to increase the range of quantum key distribution, but the application to interferometric telescopes has more stringent demands in a number of ways.  Quantum repeaters are still under development, and our work provides a new goal for research in that area.  It sets a new slate of requirements for the technology, but simultaneously broadens the appeal of successfully building quantum repeater networks.

\begin{figure}
\centering

\includegraphics[width=8cm]{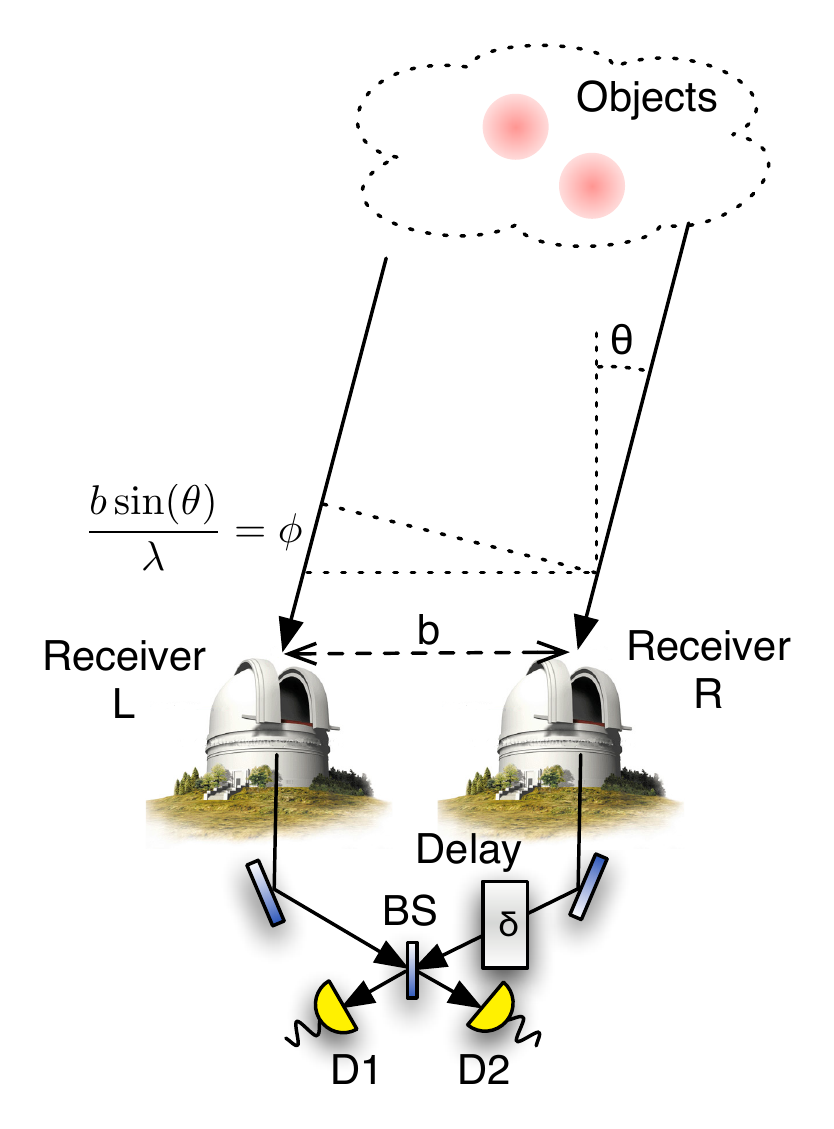}
\caption{Basic set-up of a direct detection interferometer.  In the arrangement pictured, light travels an additional distance $b \sin \theta$ to reach telescope L rather than telescope R. For light with wavelength $\lambda$, the extra distance imposes a phase shift $\phi = (b \sin \theta)/\lambda$ at telescope L relative to telescope R.}
\label{fig:directdetection}
\end{figure}

We begin by reviewing the standard approach to optical and infrared interferometry, known as ``direct detection,''~\cite{Monnier,Michelson} but we will treat the arriving light quantum-mechanically.  The light is essentially in a weak coherent state, but the average photon number per mode is much less than $1$,
% include explicit number? Mention post-selection to eliminate vacuum?
so two-photon events are negligible.  Therefore, we assume the incoming wave consists of a single photon.  We consider first an idealized set up with two telescopes and no noise, as in figure~\ref{fig:directdetection}.

Depending on the orientation of the ``baseline'' (the relative position of the telescopes in the interferometer), the light has a relative phase shift $\phi$ between the two telescopes L and R, resulting in the state:
\begin{equation}
\ket{0}_L \ket{1}_R + e^{i\phi} \ket{1}_L \ket{0}_R,
\end{equation}
with $\ket{0}$ and $\ket{1}$ indicating $0$ and $1$-photon states.
If we measure $\phi$ with high precision, that tells us the source's location very precisely.  $\phi$ is proportional to the baseline, so longer baselines produce a more accurate measurement of the source's position.

Often we are interested in sources that have structure on the scale we can resolve with the interferometer.  Different locations on an astrophysical source usually emit light incoherently, so the light is in a mixed state, formed by a mixture of photons from different locations on the source.  Because different locations give different phase shifts $\phi$, the off-diagonal components of the density matrix decrease.  We get a density matrix of the form
\begin{equation}
\rho = \frac{1}{2} \left(
\begin{array}{cccc}
0 & 0 & 0 & 0 \\
0 & 1 & \vis^* & 0 \\
0 & \vis & 1 & 0 \\
0 & 0 & 0 & 0
\end{array}
\right)
\label{eqn:visibility}
\end{equation}
in the basis $\ket{0}_L \ket{0}_R$, $\ket{0}_L \ket{1}_R$, $\ket{1}_L \ket{0}_R$, $\ket{1}_L \ket{1}_R$.
$\vis$ is known as the ``visibility.''  $\vis (\vec{b})$ is a function of the baseline vector between the telescopes.

The light from the two telescopes is then brought together.  The light from telescope R is subjected to an additional delay relative to the light from L so that when the photons are combined in the interferometer, the path travelled by an L photon differs from the path of an R photon by less than the coherence length of the incoming light. The delay line is adjustable, producing a known phase $\delta$ for the light from telescope R.  In figure~\ref{fig:directdetection}, the light then enters a beam splitter.  We see the photon in output port $1$ with probability
$\left[ 1+ {\mathrm Re}\left( \vis e^{-i \delta} \right) \right]/2$,
and in output port $2$ with probability $\left[1 - {\mathrm Re} (\vis e^{-i \delta})\right]/2$.  By sweeping through different values of $\delta$, we can measure both the amplitude and the phase of $\vis$.

A single pair of telescopes with a fixed baseline doesn't produce enough information to reconstruct the original source brightness distribution, but an array of telescopes with many different baselines acquires much more information.  The van Cittert-Zernike theorem~\cite{Zernike} states that the visibility (as a function of baseline) is the Fourier transform of the source distribution.  Thus, if we could measure the visibility for all baselines, we could completely image the source.  With only a limited number of baselines, the discrete Fourier transform may nonetheless give a good approximation of the source brightness distribution.

There are two major difficulties involved in implementing the set-up described in figure~\ref{fig:directdetection}.  First, if the telescopes are ground-based, density fluctuations in the atmosphere modify the relative phase shift between the telescopes.  The phase noise is large enough to completely swamp the signal.  Our proposal suffers from this problem just as do direct-detection interferometers, and the same solutions to it apply.  For instance, one can use space-based telescopes, perform phase referencing to recover the original phase information, or, in an array of many telescopes, calculate closure phases, which combine the interference results from different pairs of telescopes to cancel out telescope-specific phase shifts due to atmospheric fluctuations or other causes~\cite{Monnier}.

The second problem is that it is difficult to transport single photons over long distances without incurring loss of photons and additional uncontrolled phase shifts. For instance, slight variations in path length due to vibrations or small misalignments of the optical elements both produce reduced interference fringes.  The signal we wish to measure is the amount of interference --- for instance, a point source should have complete constructive and destructive interference, while a uniformly bright field of view should have no interference at all.  Since many different error mechanisms also cause a reduction in the interference visibility, this is a serious problem.  Loss of photons can present a severe limitation on the array's sensitivity to faint sources.  In practice, these problems limit the baseline size of interferometers using direct detection.  Today's best optical and infrared interferometers use baselines of only a few hundred meters at most.  This is the problem we wish to address.

The task of transporting quantum states reliably has been intensively studied in the field of quantum information.  For the specific task of interferometry, we suggest using a ``quantum repeater''~\cite{Repeaters,BDCZ}: Instead of sending a valuable quantum state directly over a noisy quantum communications channel, instead create a maximally entangled state~\cite{TWC} such as $\ket{01} + \ket{10}$, and distribute that over the channel.  The entangled state is known and replaceable, so we can check to see that it has arrived correctly.  If it has, then we transmit the original quantum state using a technique known as ``quantum teleportation''~\cite{teleportation}.

%To teleport a single-qubit quantum state (i.e., one with a two-dimensional relevant Hilbert space) from an experimenter Alice to an experimenter Bob, Alice and Bob must first share a maximally entangled state.  Then Alice performs a Bell measurement (i.e., projects on the states $\ket{00} \pm \ket{11}$, $\ket{01} \pm \ket{10}$) between the qubit she wishes to teleport and the entangled state.  The measurement result can be described as two classical bits, which Alice sends to Bob.  Bob then completes the teleportation procedure by performing a single-qubit gate which depends on the message he receives from Alice.  This reconstructs exactly Alice's original qubit state.

\begin{figure}
\centering
\includegraphics[width=8cm]{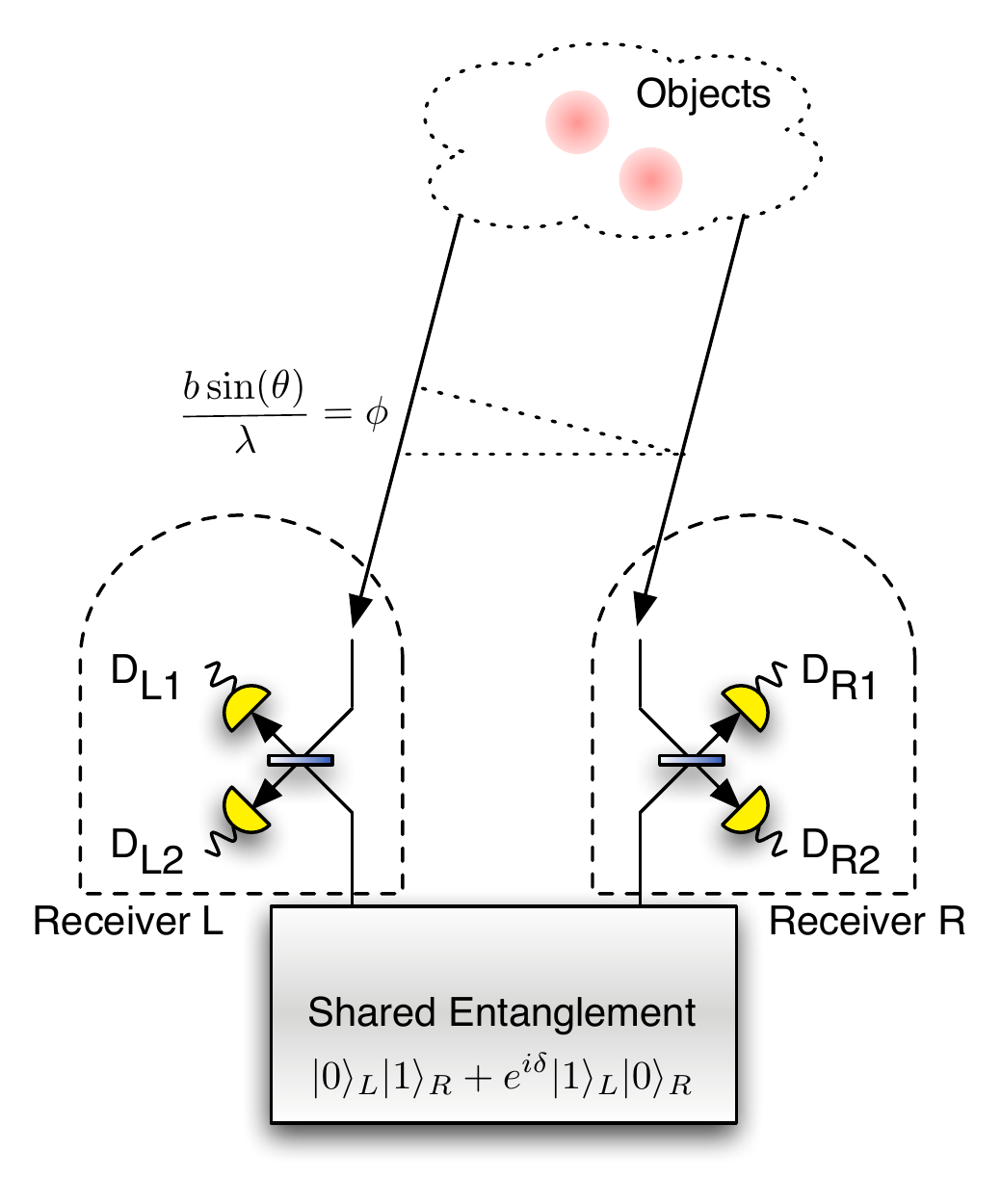}
\caption{Performing an interference measurement between two telescopes using an entangled state emitted from a central entangled photon source (EPS).}
\label{fig:quantum}
\end{figure}

For an interferometric telescope, it is not necessary to perform the teleportation explicitly; we can use the entangled pair directly to measure the visibility, as in figure~\ref{fig:quantum}.  We now have two separate interference measurements, one at each telescope.  We post-select on the measurement results, considering only the case where we see one photon at telescope L and one photon at telescope R.  One of these photons has come from the astronomical source, and one has come from the entangled pair, but we have no way of knowing which is which.  We refer to them as the ``astronomical'' photon and the ``lab'' photon, respectively.  On each side, there are two detectors, and the probability of seeing a photon at the two detectors is equal.  The signal we wish to measure is contained not in the number of photons seen at any given detector, but in the correlation between which L detector clicks and which R detector clicks.

Again, we assume the state of the astronomical photon is given by equation~(\ref{eqn:visibility}).  The variable delay line is now applied to the entangled state when the photon is sent to L, producing the entangled state $\ket{0}_L \ket{1}_R + e^{i \delta} \ket{1}_L \ket{0}_R$.  Note that the interference measurement at detector L occurs slightly later than the interference measurement at R.  When we post-select, we insist that the observed photons be displaced by precisely this time delay, with an uncertainty given by the coherence length of the photons.

Half the time, both photons arrive on the same side.  We discard those cases.  We lump together pairs of outcomes where there is one photon on each side.  The total probability of seeing a correlation (L1, R1 or L2, R2), conditioned on having one click at each telescope, is $[1 + {\mathrm Re} ( \vis e^{-i\delta})]/2$, and the total probability of seeing an anticorrelation (L1, R2 or L2, R1) is $[1 - {\mathrm Re} ( \vis e^{-i\delta})]/2$.  The measurement of correlation vs.\ anticorrelation thus provides the same information as the two outputs of the beam splitter in a direct detection experiment.

Figure~\ref{fig:quantum} can be interpreted as a post-selected teleportation at R followed by an interference experiment at L.  The beam splitter and photo-detectors at R implement a measurement with projectors $\ket{0}_A \ket{1}_E \pm \ket{1}_A \ket{0}_E$, where the subscript A denotes an astronomical photon mode and E denotes a mode of the entangled photon.  When $0$ or $2$ photons arrive at R, the teleportation fails and we discard the state, but when $1$ photon is detected at R, we succeed in teleporting the arriving A state to L, where it is interfered with the A mode arriving at L.  Of course, the diagram is completely symmetric, so we can equally well consider it as teleporting the state from L to R.

In principle, the sensitivity of an entangled-state interferometric telescope can be similar to that of a direct-detection interferometer, but there are a number of significant technological barriers to achieving the same level of sensitivity, even without a quantum repeater.  We need a high-rate true single-photon source~\cite{Buller:2010lr,Oxbarrow:2005} which puts out exactly one photon per field mode to produce the entangled states, and very fast detectors to allow a large bandwidth. Furthermore, $50\%$ of the light will be lost in the scheme of figure~\ref{fig:quantum}, corresponding to cases where the astronomical photon and entangled photon arrive at the same telescope.  The loss can be reduced to $1/n$ for an array of $n$ telescopes by using a ``W'' state as the entangled state, consisting of a single photon split coherently between the $n$ telescopes.  These and other issues relating to implementation of the scheme are discussed in more detail in the supplemental material.

Our scheme's advantage is that it allows extending the baseline of interferometers well beyond what is currently possible.  There is a substantial body of research investigating how to create entangled states shared by faraway sites~\cite{Repeaters}, and our scheme allows us to apply those techniques to the problem of creating long-baseline interferometers.

A quantum repeater can help us establish an entangled state at the two telescope locations by reducing two common types of noise.  The first challenge is phase noise, often due to path length variation in the interferometer.  Active stabilization of path lengths can substantially reduce phase noise~\cite{stabilization}.  Another solution to phase noise is entanglement distillation~\cite{distillation}, a protocol which takes a number of noisy entangled states as input and outputs a smaller number of less-noisy entangled states.  Active stabilization can be applied equally to direct-detection or entangled-state interferometry, but entanglement distillation is only available for entangled-state interferometry.  The second challenge is loss of photons, under which only a fraction of the entangled states that are sent are received.  One well-known scheme to reduce loss is due to Duan et al.~\cite{DLCZ}.  In that scheme, two atomic clouds are entangled in a ``heralded'' way, meaning we have a measurement that tells us when the entanglement has succeeded despite the loss during transmission.  We continually attempt to generate entanglement between the atomic clouds, and once we succeed, we can store it until it is needed.  We discuss repeater protocols further in the supplemental material.

\begin{figure}

\centering
\includegraphics[width=8cm]{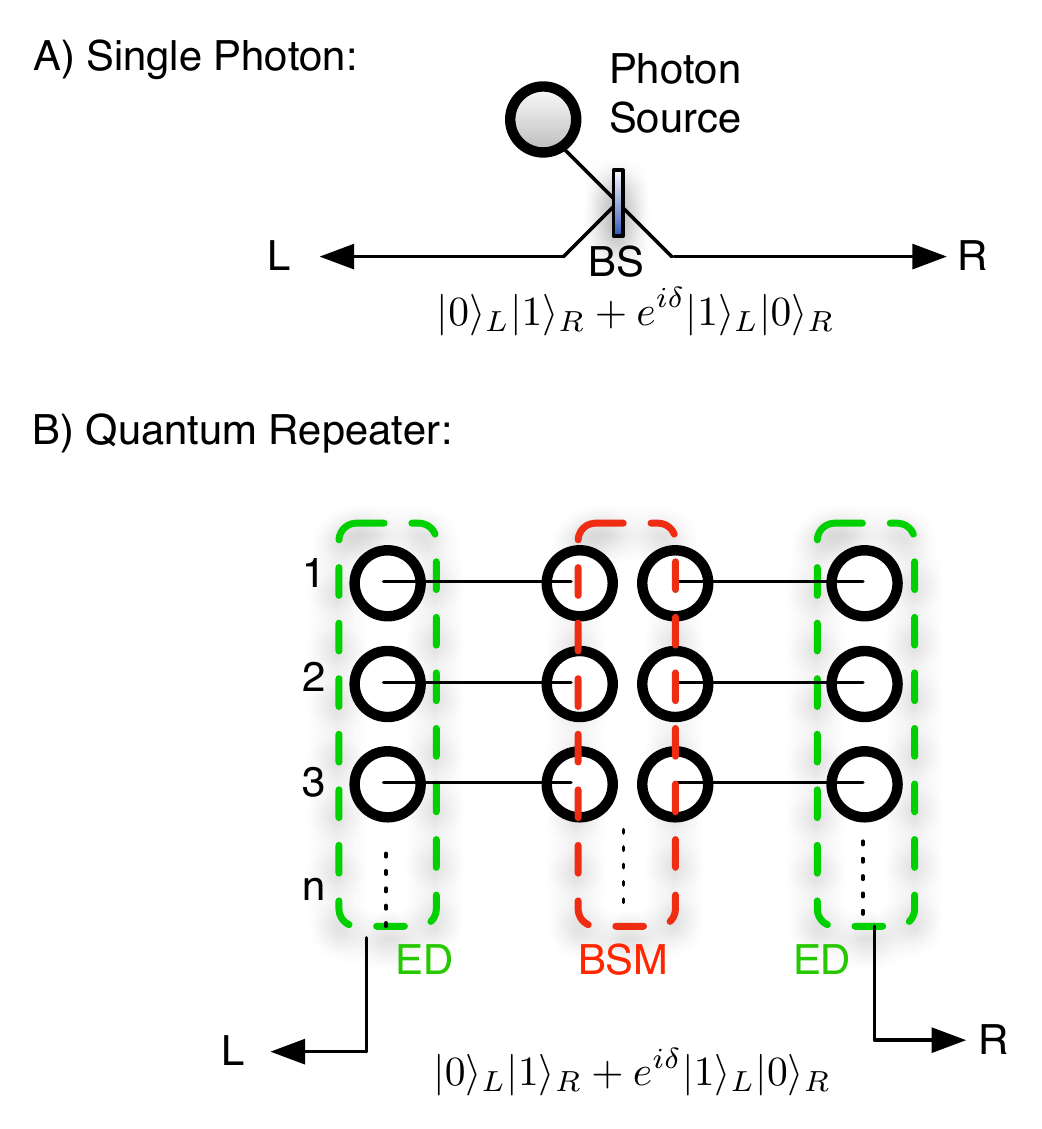}

\caption{Creating shared entanglement. A) shows the simplest scenario: pass one single photon through a beam splitter and send the resulting entangled modes to the receivers. B) In a quantum repeater, a series of quantum relays entangles several entangled photon pairs via a Bell-state measurement (BSM), and uses entanglement distillation (ED) to extract high-quality entanglement between distant receivers.}
\label{fig:repeaters}
\end{figure}

Building on the basic quantum repeater protocols, one could build a network of quantum repeaters to create entangled states shared between arbitrarily distant points~\cite{BDCZ}.  Repeater stations are positioned at a modest distance from each other, so that transmission errors and loss between neighboring stations are correctable via the repeater protocols described above.  We can create entangled pairs shared between neighboring repeaters, then join together multiple entangled states as in figure~\ref{fig:repeaters}, using entanglement swapping~\cite{swapping} to create an entangled state between any pair of nodes in the network.

Our protocol is a very demanding application of quantum repeater networks. In order to get a sense for the required photon rates and the sensitivity, we define a figure of merit $s = r p \Delta \lambda$.  $r$ is the rate of entangled states output by the repeater network, measured in entangled states per spatiotemporal mode ($0 \leq r \leq 1$).  $\Delta \lambda$ is the optical bandwidth for the system, which requires the repeaters to produce entangled states of bandwidth $\Delta \lambda$, and constrains the speed of the detectors, which must distinguish between photons arriving at that bandwidth.  $p$ is the optical transmission and detection efficiency  in the components of the system, not counting the $50\%$ inherent transmission due to post-selection. The rate of detected signal events (involving one astronomical photon and one lab photon) is equal to $s/2$ times the rate of astronomical photons per unit bandwidth hitting the telescopes, which is derived from the wavelength $\lambda$, the aperture size, and the magnitude of the star being observed.
%
% From: http://www.astro.ljmu.ac.uk/~ikb/convert-units/node1.html
%
% $f_{\nu} = 6.626 \times 10^{-8}  \lambda  f'_{\lambda}$
% $m_{\rm AB} = -2.5 \log f_{\nu} + 8.9$
% $f_{\nu}$ is the energy flux measured in ${\rm Jy}$ (Janskys, equivalent to 10-26 W m-2 Hz-1),
% $f'_{\lambda}$ is the photon flux measured in $s^{-1} m^{-2} \mu m^{-1}$,
% $\lambda$ is the wavelength measured in $\mu m}$,
% $m_{\rm AB}$ is the magnitude scale defined by Oke & Gunn (1983).
% The conversion to standard magnitudes depends on the calibration of the Vega flux in different wavelength bands.
% At 545 nm, AB mag = Vega mag; At 797 nm, AB mag = Vega mag + 0.45
% At 800 nm, mag 7.5, s=0.025 means ~100 events/sec/m^2.
%

We need to have more signal photons than dark counts, and enough photons arriving in an atmospheric fluctuation time (around $10$~ms~\cite{Monnier}) to measure the visibility.  %Additional spectral channels increase the number of photons per second but additional detectors would also increase dark count rates.
Assuming $1$~m receiver apertures,   $r = p = 0.5$ and $\Delta \lambda = 0.1$~nm at $\lambda = 800$~nm (corresponding to $1.5 \times 10^{11}$ entangled photons per polarization per second), we have  $s = 0.025$~nm.  Then the system is sensitive to stars with apparent magnitude around $7.5$.  This is comparable to the sensitivity of today's CHARA interferometer array~\cite{CHARA}, which also uses $1$~m telescopes.  Today's repeater protocols are nowhere near capable of working at this bandwidth, nor can they achieve this rate of entangled state production.  Achieving this sensitivity with $30$~km-long baselines (a hundredfold improvement over CHARA) would produce a very useful astronomical observatory.  Even a somewhat lower sensitivity with baselines of this size would in some respects be an improvement over existing instruments, with better angular resolution but lower sensitivity.

We also want the quantum repeater output to have a high fidelity to the correct entangled state.  In particular, if the quantum repeater occasionally produces two entangled states in the same mode, this leads to spurious detection events where the photons at L and R are both entangled photons.  The effect is much the same as having dark counts, so the rate of double entangled state production should be comparable to the rate of dark counts (say about $100$ per second).

Let us compare our scheme to other interferometric techniques.  Both intensity interferometry~\cite{Hanbury56b} and heterodyne interferometry~\cite{Townes,Tsang} can achieve much longer baselines than direct-detection interferometry, and they are technically much easier than entangled-state interferometry.  However, neither is sensitive enough to be generally applicable for interferometry in optical wavelengths except for the brightest sources, whereas entangled-state interferometry could be, if the technical hurdles we have discussed can be overcome.  Both schemes are related to entangled-state interferometry, and we discuss the connections in the supplemental material.

In this paper, we have primarily considered how distributed quantum entanglement can improve optical interferometry.  For radio frequencies, interferometry can be performed robustly today even between telescopes spread across the planet.  At optical frequencies, many fewer photons arrive per mode, making interferometry much more difficult.  In telescope design, the arriving light is usually treated classically, but when the number of photons arriving is small, the quantum state of the light may become important.  Thus, the field of quantum information is well-suited to provide advances.

Quantum repeaters have until now been under development primarily for use in quantum communications, so interferometry offers an interesting new venue for the application of quantum information techniques.  As we have shown, quantum repeaters can completely lift the upper limit on distance over which it is possible to do interferometry, but a number of technical hurdles need to be overcome first.  In particular, we need quantum repeater protocols capable of producing an extremely high rate of broadband entangled photons, as well as high efficiency photodetectors with fast time resolution.  One additional requirement we have is that we would like to perform astronomy at a variety of optical frequencies; either the repeater protocols need to work at those frequencies or we need a way to shift the frequencies~\cite{McGuinness} of either the arriving light or the entangled photons.

Quantum information technology may offer even further significant applications to help improve astronomical observation, even beyond direct quantum detection techniques \cite{doi:10.1080/09500340600742270}.  For instance, it may be advantageous to coherently store arriving photons using a quantum memory and then perform the quantum Fourier transform, rather than measuring and performing the classical Fourier transform.  The quantum Fourier transform works reasonably well even with a small number of photons, whereas if we measure first, we need enough photons to get a reliable measurement of each phase.

\section*{Acknowledgements}

We would like to thank Andy Boden, Latham Boyle, Avery Broderick, Jean-Philippe Bourgoin, Ignacio Cirac, Alexey Gorshkov, Liang Jiang, Jeff Kimble, Evan Meyer-Scott, Barry Sanders, and Cristoph Simon for helpful conversations.  All authors acknowledge support by NSERC, by the Government of Canada through Industry Canada, and by the Province of Ontario through the Ministry of Research \& Innovation.  D.~G.\ and T.~J.\ are supported by CIFAR, and T.~J.\ is supported by the Canadian Space Agency.
%\section*{Author Contributions}

%D~.G.\ is responsible for the main idea and many of the details of the paper.  T.~J.\ provided most of the details on detectors, photon sources, and noise, and helped brainstorm schemes.  S.~C.\ is responsible for the W state protocol and checked the quantum optics calculations.  All authors contributed to the writing of the manuscript.
%
%\section*{Author Information}
%
%The authors declare no conflicts of interest.  Correspondence should be addressed to dgottesman@perimeterinstitute.ca.

\appendix

\section*{Supplemental Material}

\subsection{Technological Difficulties in the Implementation of an Entangled-State Interferometer}

The set-up given in figure~1 of the main paper could be implemented using currently existing technology by replacing the optics in an existing telescope array.  However, the disadvantages involved in doing so outweigh the advantages, so the only reason to do so would be as a proof of principle.  In this section, we shall discuss the barriers to building an entangled-state interferometer.  For now, let us focus on the case where there is no quantum repeater, and entangled states are produced and used directly to measure the phase of astronomical photons.  The next section will address the challenges of using our protocol with quantum repeaters.

As we shall see, none of the difficulties is insurmountable.  In practice in the near term, many of these problems will reduce the sensitivity of an entangled-state interferometer relative to a direct-detection interferometer.  We shall discuss the ways these challenges can be overcome, showing that in principle, an entangled-state interferometer can have sensitivity nearly equal to that of a direct-detection interferometer.

Because we must discard the result half of the time (when both photons are on the same side), we automatically lose half of the light from the source.  If one were better able to manipulate quantum states of light, one could in principle do a full Bell measurement on each side (adding projectors $\ket{00} \pm \ket{11}$ to the existing projectors $\ket{01} \pm \ket{10}$), letting us use every arriving photon.  However, the full Bell measurement in this case is difficult to perform, so the partial Bell measurement will have to be used in near-term implementations of the scheme.  A better solution is possible when the interferometer involves an array of many telescopes.  Instead of splitting a single photon between two telescopes, let us split it between $n$ telescopes in a W state:
\begin{eqnarray}
& e^{i \delta_1} \ket{100 \ldots 0} + e^{i \delta_2} \ket{010 \ldots 0} + \nonumber \\*
& \qquad e^{i \delta_3} \ket{001 \ldots 0} + \cdots + e^{i \delta_n} \ket{000 \ldots 1}
\end{eqnarray}
The photon arriving at telescope $i$ is subject to a phase delay $\delta_i$ so that modes emitted simultaneously from the astronomical source can be compared at any pair of telescopes.  At each telescope, we put the two photon modes through a beam splitter as before, and post-select on getting two simultaneous photons at any two telescopes.  In the ideal case, one will be an astronomical photon and one will be an entangled photon.  The only cases we must discard are when both photons arrive at the same site.  This happens with probability $1/n$, so as the number of telescopes increases, we lose less and less signal from the partial Bell measurement.  Once we post-select on a particular pair of telescopes, we look for correlation or anticorrelation between the measurements at those two telescopes; the analysis is just the same as if we had originally planned an interference measurement between just those two telescopes.

The W state technique has some additional advantages and drawbacks relative to using entanglement between pairs of telescopes.  An advantage is that phase errors in the W state are telescope-specific phase errors which can be cancelled via closure phases, whereas phase errors in a $2$-telescope entangled state can be baseline-specific.  One minor drawback is that we don't control which baseline gets used for this astronomical photon.  That is not important, luckily, since to take full advantage of an array, we would want measurements on all possible baselines in any case.  Each baseline gets the same average number of astronomical photons as when we use pairwise entanglement, but we keep a fraction $1-1/n$ of them instead of $1/2$.  A more important disadvantage is that it is more challenging to combine the W state with quantum repeaters.  The W state scheme is most useful when we lack complete Bell measurements, but then we also cannot reliably perform teleportation, so the chance of successfully distributing an $n$-qubit W state over standard quantum repeaters decreases as $2^{-n}$.  There are various solutions to this.  We could use a non-standard repeater, or could build the W state up from smaller pieces.  In the absence of other imperfections, either of these can be done with an amount of resources which is polynomial in $n$.

Because we want a separate entangled state for each photon mode, we will need to be able to produce entangled pairs at a very high rate.  In theory, this can be substantially reduced: Since most of the astronomical photon modes are empty, quantum compression~\cite{Schumacher} applied to those modes can greatly reduce the number of qubits that need to be teleported.  When a universal set of quantum gates is available, this can be done with only a small amount of scratch space as photons arrive~\cite{BC,BCG}.  Indeed, it may be possible to reduce the amount of entanglement even further since we don't require actually sending the state; we only want to know the relative phase at telescope L vs.\ R.  However, it doesn't appear to be possible to do the compression via only linear optics, so it is an interesting theoretical question whether there is some way of compressing the states without excessive experimental difficulty.

There is also a loss of signal caused by imperfect detector efficiency.  A standard direct-detection interferometer requires a click from only one photodetector, but our scheme requires two photodetectors to click.  Therefore, the overall signal in our scheme is suppressed by an additional factor of the detector efficiency.  In principle, detector efficiencies can be very close to $1$, but in practice they are not.  Further development of high-efficiency detectors will therefore help make our scheme more practical.

We have described the experiment assuming there is exactly one photon from the astronomical source and one photon in our entangled state.  In practice, neither may be true.  The easiest ``entangled'' state to create would use a weak coherent state, such as an attenuated laser, as a photon source.  The light from the astronomical source has undergone very severe attenuation since its emission, so it is also best modeled as a weak coherent state.  Assume that the ``entangled state'' has average photon number $p_E$ and the astronomical source has average photon number $p_A$.  Both $p_E$ and $p_A$ are substantially less than $1$.  Splitting up a weak coherent state gives two weak coherent states, with average photon numbers $p_E/2$ for the entangled state and $p_A/2$ for the astronomical source.

We post-select on seeing one photon at each telescope, and those events come from three cases.  We will keep only the lowest-order ($p^2$) terms, under the assumption that both $p_E, p_A \ll 1$.  In the first case, with probability $p_A p_E/2$ we have one lab photon and one astronomical photon, and then we see an interference pattern in the correlation between which L detector and which R detector clicks.  The next case, which occurs with probability $p_E^2/4$, is when there are two lab photons, and then the left and right detector outcomes are uncorrelated.  Finally, in the third case, there are two astronomical photons, and again the L and R detections are uncorrelated.  The Hanbury Brown-Twiss effect~\cite{Hanbury56b} comes into play, so the probability of this case is $p_A^2(1 + {\mathrm Re} \vis)/4$.
The total probability of getting one photon at L and one photon at R is
\begin{equation}
P = \frac{p_A p_E}{2} + \frac{p_E^2}{4} + \frac{p_A^2}{4} (1 + {\mathrm Re} \vis).
\end{equation}
For the $p_A^2$ and $p_E^2$ terms, both photons come from the same place, and the left and right detector outcomes are uncorrelated.  Therefore, when the source is a weak coherent state, the probability of seeing a correlation between the L and R detectors is
\begin{equation}
\frac{1}{8} \left[ p_E^2 + p_A^2 (1 + {\mathrm Re} \vis) + 2p_A p_E + 2 p_A p_E {\mathrm Re} \left( \vis e^{-i\delta} \right) \right].
\end{equation}
Normalizing by the total number of events, the visibility term, where our signal resides, is decreased by a factor
\begin{equation}
\frac{2 p_A p_E}{p_E^2 + p_A^2 (1 + {\mathrm Re} \vis) + 2p_A p_E}.
\label{eqn:coherentvisibility}
\end{equation}
One might be puzzled why we see any interference at all, since the purported entangled state is actually a tensor product of two weak coherent states. The answer is that we are dealing with a post-selected measurement.  Conditioned on seeing one photon on each side, we actually do have an entangled state.

The source strength $p_A$ is fixed, but we can vary $p_E$ to our liking to get the best result.  The visibility loss is minimized, according to formula~(\ref{eqn:coherentvisibility}), when $p_E \approx p_A$.  However, choosing that value of $p_E$ means that we rarely see the two-photon events we are interested in.  An alternative choice is to let $p_E$ be larger to increase the rate of two-photon events, but we then have to suffer the concomitant loss of normalized visibility.  This is an unappealing tradeoff: using a weak coherent state source, we must choose between losing most of the light that arrives or reducing our signal-to-noise ratio.  A better solution is to use a true single-photon source, which reliably emits a single photon when asked and never emits two photons at the same time~\cite{Buller:2010lr,Oxbarrow:2005}.  Such sources are under development, so this is not an unreasonable requirement.  However, we need a single-photon source which produces a photon indistinguishable from the astronomical photon.  Furthermore, the reset time between emissions of a photon leads to loss of signal, as any astronomical photon which arrives during the dead time is not measured.  In other words, we want the single-photon source to produce an entangled photon for every astronomical photon mode we are measuring.  That is much more challenging, and should become a goal of research into single-photon sources.

Another substantial challenge is that the entangled photon and astronomical photon must be nearly indistinguishable to produce good interference at the beam-splitters.  Ideally, the spatial mode of the astronomical photon is known (essentially a plane wave truncated to the size and shape of the telescope aperture), and the frequency mode is controlled by a spectral filter placed in the telescope.  In practice, the spatial mode will be distorted by atmospheric effects, but this is also a problem for direct-detection interferometers, and can be dealt with using the same methods (e.g., adaptive optics or single-mode fibers)~\cite{Monnier,Michelson}.  The main challenge is matching the temporal mode correctly.  The photons will be distinguishable if they arrive separated by a time greater than their coherence time.  Usually, the astronomical source being observed is very hot, so the light emitted has a large bandwidth and correspondingly low coherence time.  The spectral filter narrows the bandwidth and therefore increases the coherence time, according to the uncertainty principle.  Our choice of what bandwidth to allow therefore sets the requirement for time resolution.

In particular, if the detectors in use have a time resolution greater than the coherence time, many supposedly simultaneous two-photon coincidences will actually be between distinguishable pairs of photons, which will reduce the interference fringe visibility~\cite{Zukowski95a}.  Currently, the lowest timing resolution achieved is $\tau_c = 35$ ps (FWHM) with thin Silicon Avalanche Photo Diodes~\cite{Buller:2010lr}, with a high detection efficiency of above $50\%$ in the green ($\lambda = 550$ nm). Consequently, in order to achieve a high two-photon interference contrast the bandwidths of both the incoming and the entangled photons $\Delta \lambda$ must be matched and narrowed to a value on the order of $\Delta \lambda = \frac{\lambda^2}{2 \pi \tau_c } \approx 0.025$ nm (FWHM), truly challenging but not impossible.  Furthermore, it is in principle possible to split out the spectrum into several wavelength channels which can each be measured separately in order to enhance the sensitivity of the system.  Relative to a single broad bandwidth channel, many narrow channels have other advantages:  For instance, they give us additional information about the frequency dependence of the source and loosen the restriction on field of view caused by baseline smearing, which can be substantial at very large baselines~\cite{Monnier}.

The ultimate limit on the sensitivity of the receivers will be determined by the noise dark counts in the photon detectors (ca. 100 counts per second), which must be less than the number of original photons received. In addition, the number of astronomical photons received must be significant to perform a visibility measurement within the characteristic time of the atmospheric phase fluctuations, which is on the order of 10~ms~\cite{Monnier}. Assuming a 1~m receiver aperture, this threshold should roughly be surpassed when observing objects of an apparent magnitude of 7.5.  This is comparable to the sensitivity of today's CHARA interferometer array~\cite{CHARA}, which also uses 1~m telescopes.  Of course other imperfections in the system will hurt our sensitivity further, but the point of the quantum repeater protocols, discussed below, is that losses and noise due to the transmission of the photons can be largely eliminated, so it is reasonable to expect that we can attain a sensitivity close to the ideal.  With the ongoing advancement of photon detector technology, we can imagine that timing resolutions of $< 5$ ps and dark counts of perhaps 10 cps are feasible, which would allow the bandwidth of the photons to be widened and the limiting sensitivity of the system improved to about magnitude 12.

\subsection{Quantum Repeaters and Their Application to Entangled-State Interferometry}

There are two primary sources of noise where a quantum repeater can help us.  The first problem is phase noise, often due to path length variation in the interferometer.  Active stabilization of path lengths can substantially reduce phase noise~\cite{stabilization}.  Another solution to phase noise is entanglement distillation~\cite{distillation}, a protocol which takes a number of noisy entangled states as input and outputs a smaller number of less-noisy entangled states.  Entanglement distillation and active stabilization are complementary procedures, as active stabilization is most effective against slowly changing sources of phase noise, whereas entanglement distillation works best to eliminate noise sources that are uncorrelated between the transmitted entangled states.  Active stabilization can be applied equally to direct-detection or entangled-state interferometry, but entanglement distillation is only available for entangled-state interferometry.

For instance, Sangouard et al.~\cite{SSCG} (SSCG) presented a distillation protocol specialized for single-photon entangled states of the sort we use above.  One starts with two entangled photons.  At the L end of the state, the two modes are sent through an unbalanced beam splitter, and similarly at the R end.  One output of each beam splitter is monitored with a photodetector, and if exactly one of the two detectors sees a photon, the output of the other two ports is kept as the new more reliable entangled state.  If both detectors see a photon, there are no photons left, and the state must be discarded; if neither detector sees a photon, the state has become a two-photon state, and should also be discarded.  The optimal beam splitters to use depend on the level of noise in the transmission channel, but using one $15/85$ beam splitter and one $85/15$ beam splitter is close to optimal for all noise rates.

When the repeater succeeds, it outputs an entangled state with a higher fidelity to the desired state than the original entangled states produced by the channel.  If the fidelity is still not high enough, we can repeat the protocol using two entangled states, each of which is itself the output of the first-round repeater protocol.  If that is still not good enough, we can continue for more rounds until the entanglement reaches the desired fidelity.

The SSCG protocol is not very efficient, since a majority of the entangled states are lost even if there is little noise, but it does have the advantage of being straightforward to implement.  Indeed, it has already been demonstrated experimentally~\cite{SLS+}.  One important advantage of the SSCG protocol is that it does not require any shared local oscillator between the two ends in order to perform the protocol --- the photons used in the protocol act as a phase reference for each other.  Since it is difficult to reliably share a local oscillator over a long distance, that is a big advantage.

The second problem is loss of photons; in the presence of loss, not every entangled state sent is received.  One well-known scheme to solve this problem is due to Duan et al.~\cite{DLCZ} (DLCZ).  In that scheme, two atomic clouds are entangled in a ``heralded'' way, meaning we have a measurement that tells us when the entanglement has succeeded despite the loss during transmission.  We continually attempt to generate entanglement between the atomic clouds, and once we succeed, we can store it until it is needed.  The information is stored in collective excitations of the atomic cloud, which can couple strongly to light.  This makes it possible to emit the stored entangled state as an entangled photon of the kind we want to use in our interferometer scheme.  The DLCZ scheme has also been implemented experimentally~\cite{Chou}.  Unfortunately for our purposes, the atomic clouds used in the DLCZ scheme only interact with light within a narrow bandwidth.  This is a big drawback, and we would want schemes that work for larger bandwidth; that should be a goal for future development.  We also need a very rapid repeater protocol, able to output one reliable entangled state into each optical mode, as discussed above.  Current repeater protocols are too slow, so this is another task for future research.

Even more advanced protocols are possible.  Ideally one would be able to store the astronomical photon as a qubit in a quantum computer.  Entanglement would also be generated, distilled, and stored in the quantum computer using any of a variety of protocols, some of which have much better performance than the technologically easier repeater protocols cited above.  In addition, performing the interference in a quantum computer would let us do a complete Bell measurement, avoiding the $50\%$ signal loss due to the limitations of linear optics for two telescopes, and would let us compress the astronomical signal state to more efficiently use entangled states. We are not aware of any proposed protocol that could store received light in a quantum computer with high efficiency and fidelity; a workable one would be very interesting.

There is one particular point that requires caution here.  Many of these more advanced techniques will require local oscillators, and as we have mentioned, it is difficult to arrange that the oscillators in distant locations will agree.  Active stabilization of phases could allow this, but even without that technology, all is not lost.  While it is difficult to have local oscillators at both ends which have the same phase and frequency, it is much easier to have separate local oscillators which share the same frequency but not the same phase.  For instance, we can have two lasers tuned to the same atomic line.  In this case, the repeater protocol will generally insert an additional phase into each state equal to the relative phase between the local oscillators.  This phase is unknown, but is stable over time, to the extent that the local oscillators are stable.  With an appropriate geometry of the repeater network, the unknown phase is a local phase shift at each telescope in the array, and can be eliminated by calculating the closure phases.

For the purposes of our discussion, we have focused our analysis on a particular protocol and some small variations of it.  However, the field of quantum information offers a much broader spectrum of techniques aimed at transmitting quantum states through a noisy channel.  For instance, quantum teleportation can be performed using a two-mode squeezed state by treating the field quadratures as continuous variables to be teleported~\cite{BK,contteleport,broadteleport}.  This kind of teleportation offers a full Bell measurement, unlike teleportation based on a single-qubit entangled state.  However, quantum repeaters for continuous-variable protocols are much more technically involved~\cite{GKP}, so overall the protocol we presented seems better at this time.  Another possible variation would be to use direct-detection interferometry, but to encode the quantum states into a quantum error-correcting code~\cite{QECC} to deal with errors and loss in transmission.  The procedure is probably more challenging, however, since quantum error-correcting codes generally require more resources and better quantum gates than repeater protocols.

Once the basic quantum repeater protocols are perfected, it becomes possible to build a network of quantum repeaters to create entangled states shared between arbitrarily distant points~\cite{BDCZ}.  Repeater stations are positioned at some modest distance from each other, so that transmission errors and loss between neighboring stations are correctable via the repeater protocols described above.  We can create entangled pairs shared between neighboring repeaters, then join together multiple entangled states using entanglement swapping~\cite{swapping} to create an entangled pair between any pair of nodes in the network.

One alternative that might be easier would be to use just a single repeater node located on a satellite, which would communicate directly to every telescope in the array.  Such an arrangement has been previously investigated to cryptographically link faraway sites via quantum key distribution.  The drawback is that many photons are lost on the way to the ground. Typical transmission success for a satellite-ground link of optical photons is 0.01 \cite{ISI:000234511800002}.  Even if the original source were a true single-photon source, by the time it reaches the ground, we have something very close to a coherent state.  To make this work, we would thus need to use the DLCZ protocol or some other method of correcting for loss.  Stabilizing the satellite so that the path length to the ground remains under control during an atmospheric fluctuation time is an additional daunting technical challenge, but perhaps not an insurmountable one.

In general, the technical considerations for an interferometric telescope with quantum repeaters are similar to those for the entangled-state interferometer without a quantum repeater, only now the requirements on the photon source are replaced by requirements on the output of the repeater network.  Again, we require that the repeater protocol should only produce one entangled state per mode, and once more we want it to produce one for essentially every mode we are measuring.  Whereas before, the most stringent limitation on the bandwidth came from the detectors, today's quantum repeater protocols do not naturally use high-bandwidth photons.  We therefore suggest developing repeater protocols capable of dealing with larger bandwidths should be a priority.  Modifying the bandwidth and/or frequency of the light might be one solution~\cite{Kielpinski,McGuinness}.

For quantum key distribution, the added technical complexity of repeaters is only justified over distances for which loss is prohibitive, i.e., hundreds or thousands of kilometers.  For interferometric telescopes, even moderate amounts of loss should be avoided as much as possible.  Therefore, our protocol can potentially be better than direct-detection interferometry over much shorter distances, perhaps $10$~km or less.  Our demands on repeaters are more stringent than for quantum key distribution in a number of ways, but in this one respect, it may be easier to build repeaters for entangled-state interferometry.  Shorter distance repeaters will have higher success rates, since the loss is less, and require shorter storage times for entangled pairs, which is technically easier.

\subsection{Comparison With Heterodyne and Intensity Interferometry}

It is instructive to compare our quantum repeater-based interferometer with some other kinds of interferometry.  In intensity interferometry, also known as the Hanbury Brown Twiss effect~\cite{Hanbury56b}, two photons from different locations in the source arrive at different telescopes.  The interference occurs between the cases where we switch which photon goes to which telescope.  Mathematically, our quantum repeater protocol is very similar to intensity interferometry, except that one of the photons is coming not from the astronomical source but is instead under our control.  That has various advantages: In intensity interferometry, we need two photons to arrive at almost the same time from a source that is not very bright; using a quantum repeater, we can in principle use a single-photon source that always outputs a photon when we need one.  Therefore, in principle our quantum repeater can be nearly as sensitive as a direct detection experiment, whereas intensity interferometry is necessarily much less sensitive.  Also, since the entangled state is under our control and comes from a different direction, we can delay it and use beam splitters to measure its phase relative to that from the source, allowing a complete measurement of the complex visibility, whereas intensity interferometry usually loses some information.  Of course, intensity interferometry has some big advantages too, namely that it is technically much less demanding than a quantum repeater protocol.

In heterodyne interferometry, light coming in to each telescope in the array is mixed via beam splitter with a laser, and photodetectors measure the relative phase between the photon from the source and the laser~\cite{Townes}.  In order to make full use of this information, the lasers at different locations should be phase-locked.  The usual way to assure this is to start with just one laser and split it up, sending the beam to different locations.  The resulting set up looks very much like our quantum repeater set up.  There are two differences: the classical processing done is different, and in heterodyne interferometry, the laser connecting different telescopes is strong, whereas our entangled state is very weak, involving at most one photon.

These differences have important consequences.  First, the bandwidth for heterodyne interferometry is determined by the speed of the electronics, and may be even narrower than we need for our entangled-state interferometry protocol.  Secondly, in heterodyne interferometry, there is no entanglement between the two telescopes, whereas our protocol relies on it.  Heterodyne interferometry is ultimately limited by quantum noise in the separate measurements at L and R, which can swamp the signal we are trying to see when the source is very faint.  For most astronomy in optical frequencies, the occupation number of each mode is much less than $1$, so heterodyne interferometry has too poor a signal-to-noise ratio to be useful~\cite{Townes,Tsang}.  That problem does not afflict our quantum repeater protocol, since the entanglement between the telescopes lets us compare correlations in the measurement outcomes.  The quantum noise is not gone --- it appears in the fact that the measurement on each side (L or R) is, by itself, completely random.  The use of an entangled state means that the noise on the L and R sides is correlated, so it cancels out when we look at the correlation/anti-correlation between the measurements.

\end{document}